\documentclass[11pt]{article}
\usepackage[T1]{fontenc}
\usepackage[utf8]{inputenc}
\usepackage{lmodern}
\usepackage[margin=1in]{geometry}
\usepackage{graphicx}
\usepackage{yfonts}
\usepackage{subcaption}
\usepackage{caption}

\usepackage{booktabs}
\usepackage{float}
\usepackage{bm}
\usepackage{amsmath,amssymb,amsthm,mathrsfs}
\usepackage[authoryear,round]{natbib}
\usepackage[colorlinks=true,linkcolor=blue,citecolor=blue,urlcolor=blue]{hyperref}

\title{Welcome to the Statverse:  A Metaverse for Data Science}
\author{Ronny Vallejos,$^1$  Miguel de Carvalho,$^2$  Roberto Cruz,$^1$  Nicolás Iribarra,$^1$ \\ José Allende,$^1$ Edmundo Casas,$^3$ Francisco Marshall,$^3$ Sebastián Suárez,$^3$\\ Leopoldo Cárdenas,$^3$ Ozan Evkaya$^2$
\\  \\ 
$^1$Universidad Técnica Federico Santa María, Chile\\  
$^2$The University of Edinburgh, UK\\
$^3$Kauel Inc., Houston, TX, 77027, USA
}

\date{ 05/06/2026}

\begin{document}
\maketitle

\begin{abstract}
    This paper introduces the Statverse, a Metaverse framework designed to revolutionize the landscape of Statistical Education in the digital age. Our key goal is to report our progress and encourage others to integrate similar strategies into their programs. The proposed framework seamlessly integrates the physical and digital realms to provide an immersive environment for the nuanced representation of complex statistical concepts.  Finally, we discuss the potential impact of Statverse on advancing Statistical Education, offering a transformative approach to teaching and learning in the digital age. Statverse is the outcome of an academic partnership between  Universidad Técnica Federico Santa María (\textsc{utfsm}) and the University of Edinburgh (\textsc{uoe}). 

\end{abstract}

\section{Introduction }\label{sec:intro}
A variety of trendsetters---such as Apple, Meta, and Microsoft---are heavily investing in the Metaverse\footnote{The term Metaverse was first coined in Neal Stephenson's 1992 science fiction novel, \emph{Snow Crash}, which is centered around a virtual reality world where individuals compete for social status through the control of their digital avatars.} and Spatial Computing, in what promises to be a breakthrough for our digital and physical realities. While the field of Statistics and Data Science stands to gain significantly from these developments, discussion on leveraging them has been limited---especially in the context of Data Science Education. 

Considering this backdrop, this paper presents the Statverse---a Metaverse framework poised to transform statistical education for the digital era. Data Science Education is undergoing various transformations, as evidenced by the American Statistical Association's recent rebrand of the \textit{Journal of Statistics and Data Science Education}---previously known as \textit{Journal of Statistics Education}. Many of these changes are driven by the need to develop practical problem-solving skills \citep[e.g.,][]{hicks2018, sharifi2023}. In addition to these key issues, there is a pressing need to integrate modern spatial computing technologies into Data Science curricula. 

\begin{figure}[H]
\centering
\includegraphics[scale=0.32]{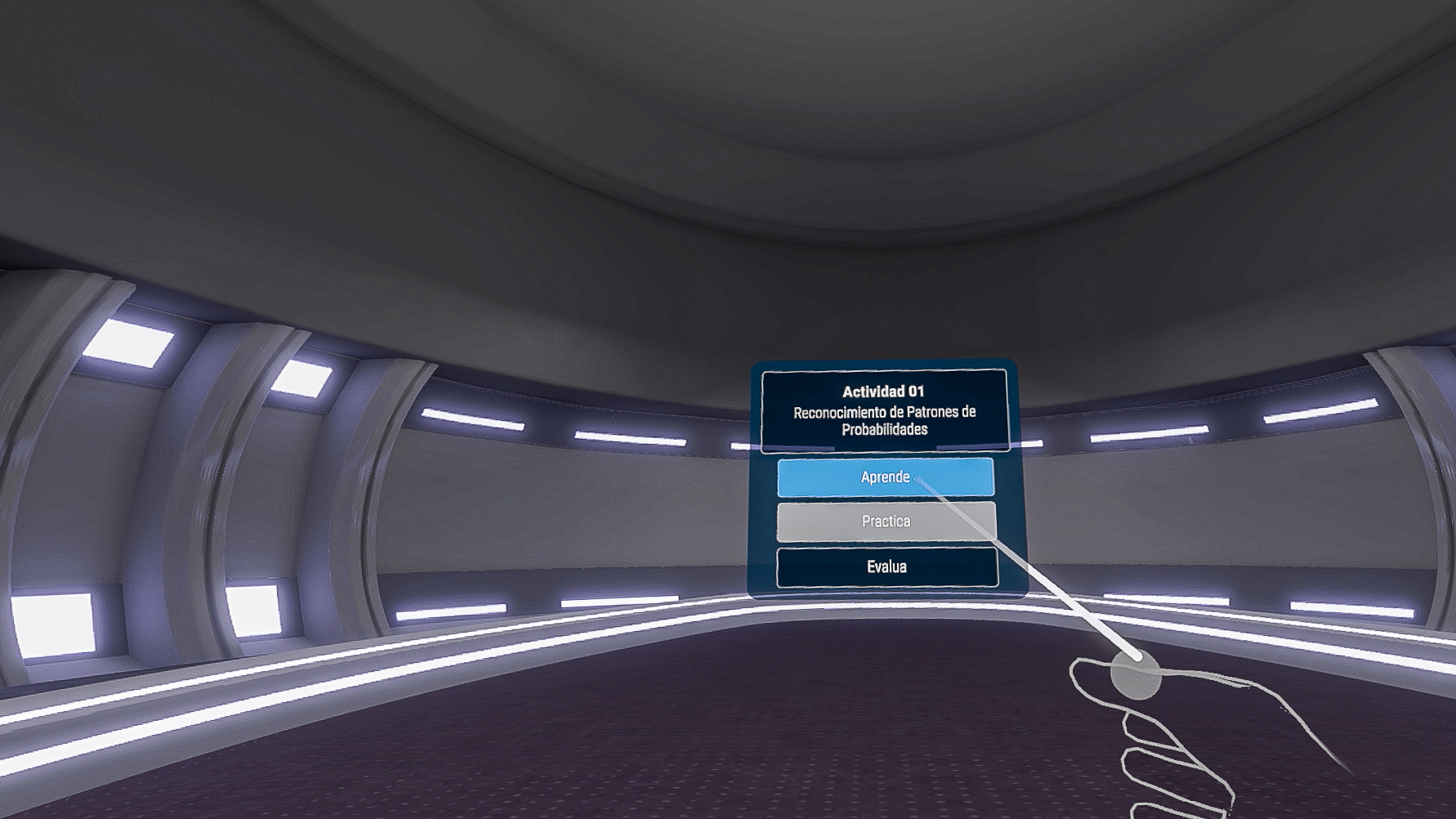}\hspace{0.01cm}
\includegraphics[scale=0.314]{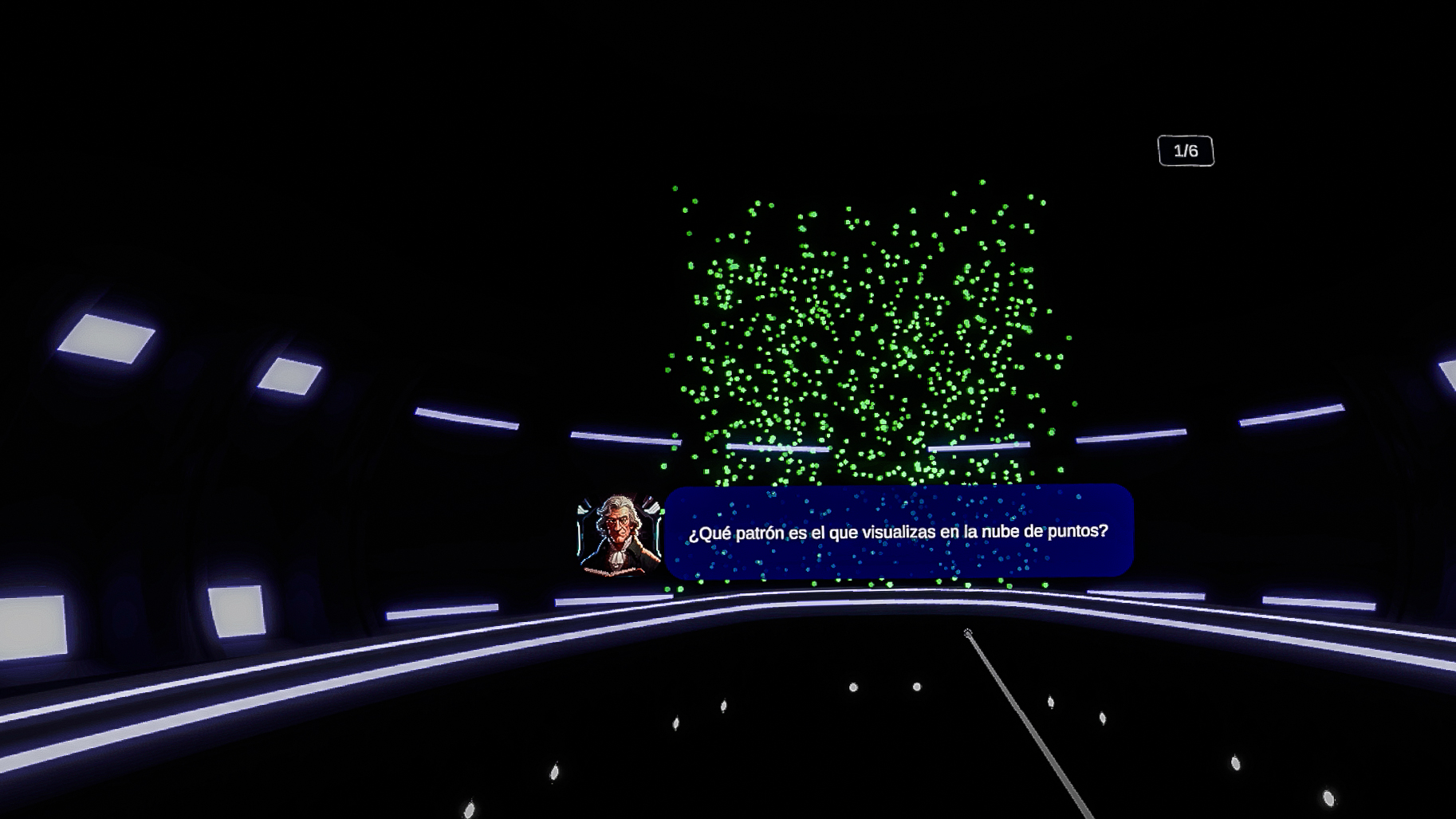}
\includegraphics[scale=0.3]{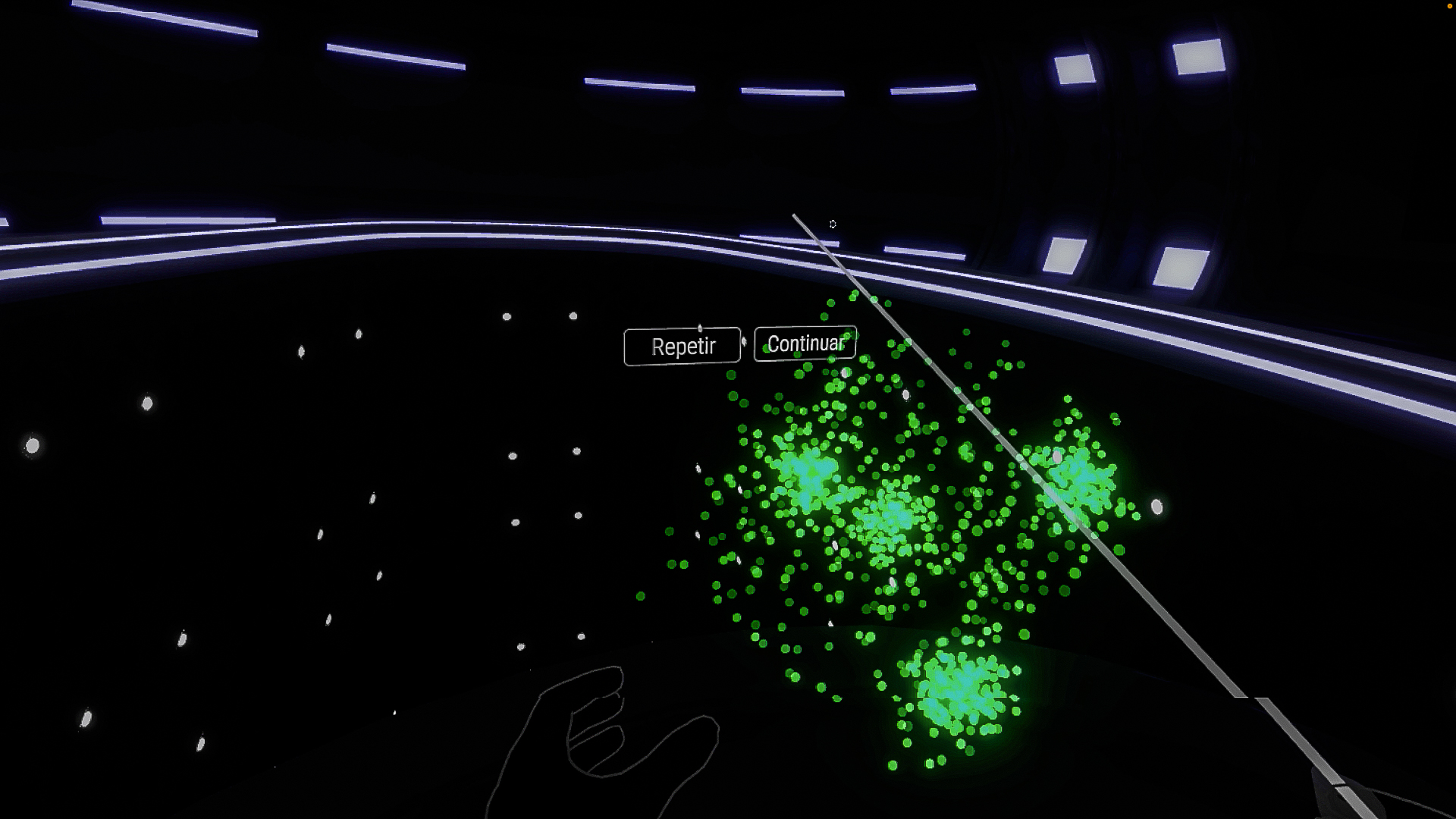}
\includegraphics[scale=0.3]{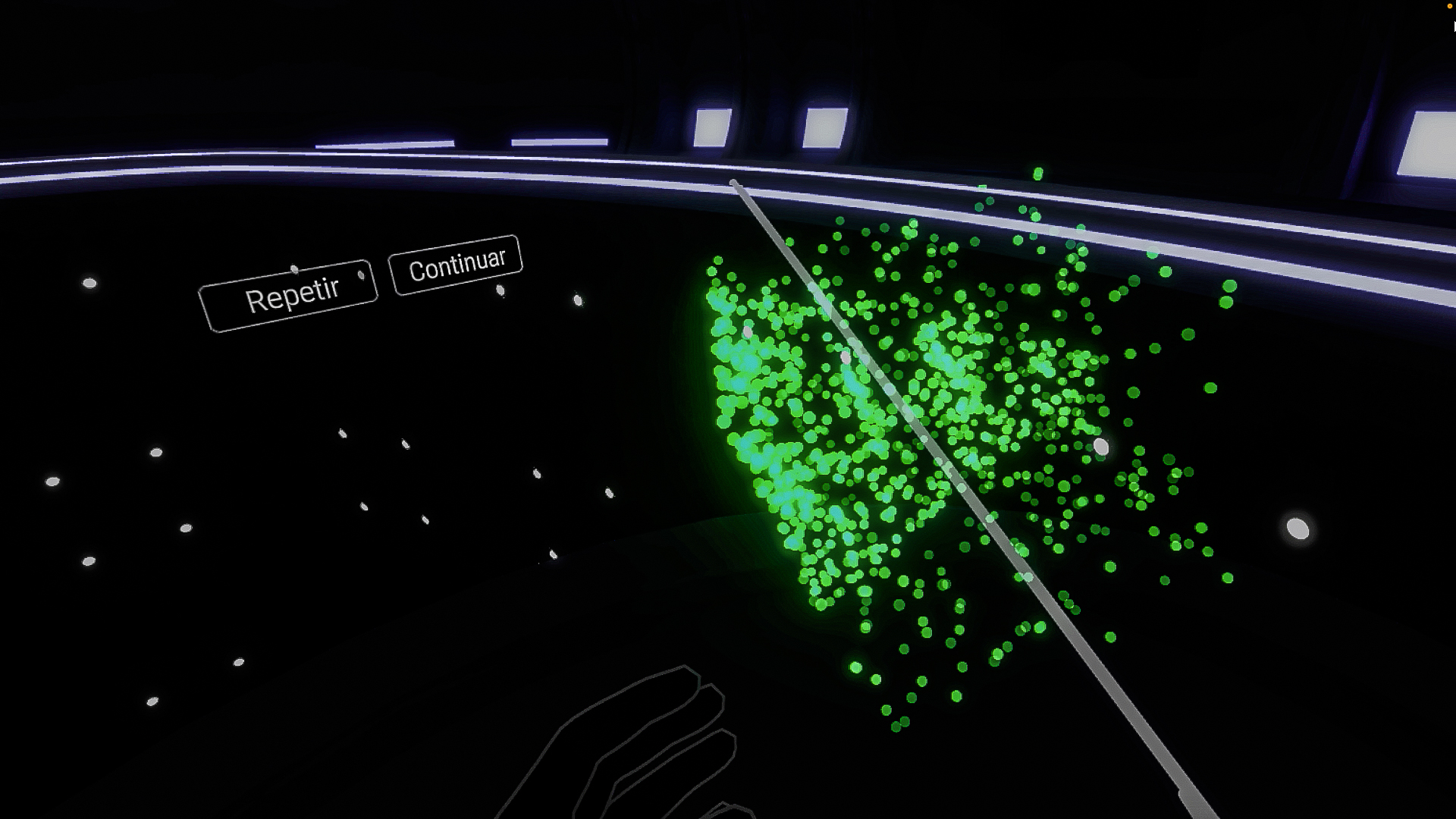}
\caption{(top left) Virtual environment of experience 1: point patterns. The rest of the pictures are snapshots from the activity in which the students have to recognize the nature of certain patterns in a 3D space.}
\label{Fig:patterns}
\end{figure}

A key objective of this paper is to document our developments in on Statverse and encourage others to adopt similar approaches in their programs. Statverse  modules, like the ones described below, offer an immersive learning experience with a variety of other dimensions that can complement those of lectures and stats labs (i.e., workshops). In Statverse, students can visualize data, engage with it, and immerse themselves in a dynamic, 3D environment where complex information becomes tangible---facilitating unparalleled learning and analysis experiences. Just as \texttt{R} shiny apps have facilitated hands-on learning and brought complex data analyses to the classroom \citep[][]{fawcett2018, wang2021}, Statverse aims to take this engagement a step further. Links for recordings that illustrate our implementation of the concept are available from: 
\begin{center}
  \url{https://edin.ac/4dIdBpD}
\end{center}
Fig.~\ref{Fig:patterns} illustrates how students can interact with a point pattern in Statverse to facilitate learning involved concepts from point processes theory.  
Beyond point processes, many other areas may benefit from the possibility of immersive real-time interaction with data, such as regression analysis, spatial statistics, and time series---just to name a few.  

\noindent As can be seen from Fig.~\ref{Fig:patterns}, and the videos above, Statverse introduces a revolutionary dimension to exploratory data analysis by enabling real-time interaction with data. It aims to provide an immersive learning experience through hands-on activities that would be impossible to replicate in a traditional classroom setting. Our framework recognizes the importance of engaging multiple human senses for a comprehensive learning experience; it aims to involve both \emph{vision} and \emph{touch} in the process of exploring data. While a lot has been written on \emph{data visualization} \citep[e.g.,][]{healy2018, wilke2019, grant2018}, much less attention has been paid to the significance of \emph{touch} for learning from data. We introduce the term \emph{data tactilization} to refer to this under-explored avenue of engaging with data through tactile experiences. Data tactilization should not be confused with tactile graphics \citep{edman1992, godfrey2012}. Tactile graphics focus on accessibility for the visually impaired, while data tactilization involves using touch to enhance interactivity with data in a metaverse environment. Thus, data tactilization broadens data visualization by incorporating multisensory engagement.

\section{Statverse unveiled}\label{unveiled}
Prior to full-scale implementation, a preliminary pilot study was conducted in the \textsc{utfsm} Admission Fair in 10/2023. The aim was to identify any technical or methodological issues.

To gain deeper insights into the development of the Statverse experiences, a brief satisfaction survey was conducted at \textsc{utfsm}. A mobile laboratory was set up and students were invited to participate voluntarily in Experience 1, aimed at exploring random patterns in space. The survey was  designed and prepared in advance to be administered upon completion of the experience, following the use of RV glasses. Using a 1 to 5 Likert scale, the survey measured the level of satisfaction and it was based on the following questions:

\begin{itemize}
\item \textbf{Question 1}: Following your participation in the Statverse activity, how would you rate your understanding of the patterns presented? \vspace{0.2cm}
\item \textbf{Question 2}: How would you rate the overall quality of the learning experience? \vspace{0.2cm}
\item \textbf{Question 3}: In your opinion, how effective is Statsverse as a teaching tool for Statistics compared to traditional instructional methods? 
\end{itemize}

\begin{figure}[htp]
\centering
\includegraphics[scale=0.28]{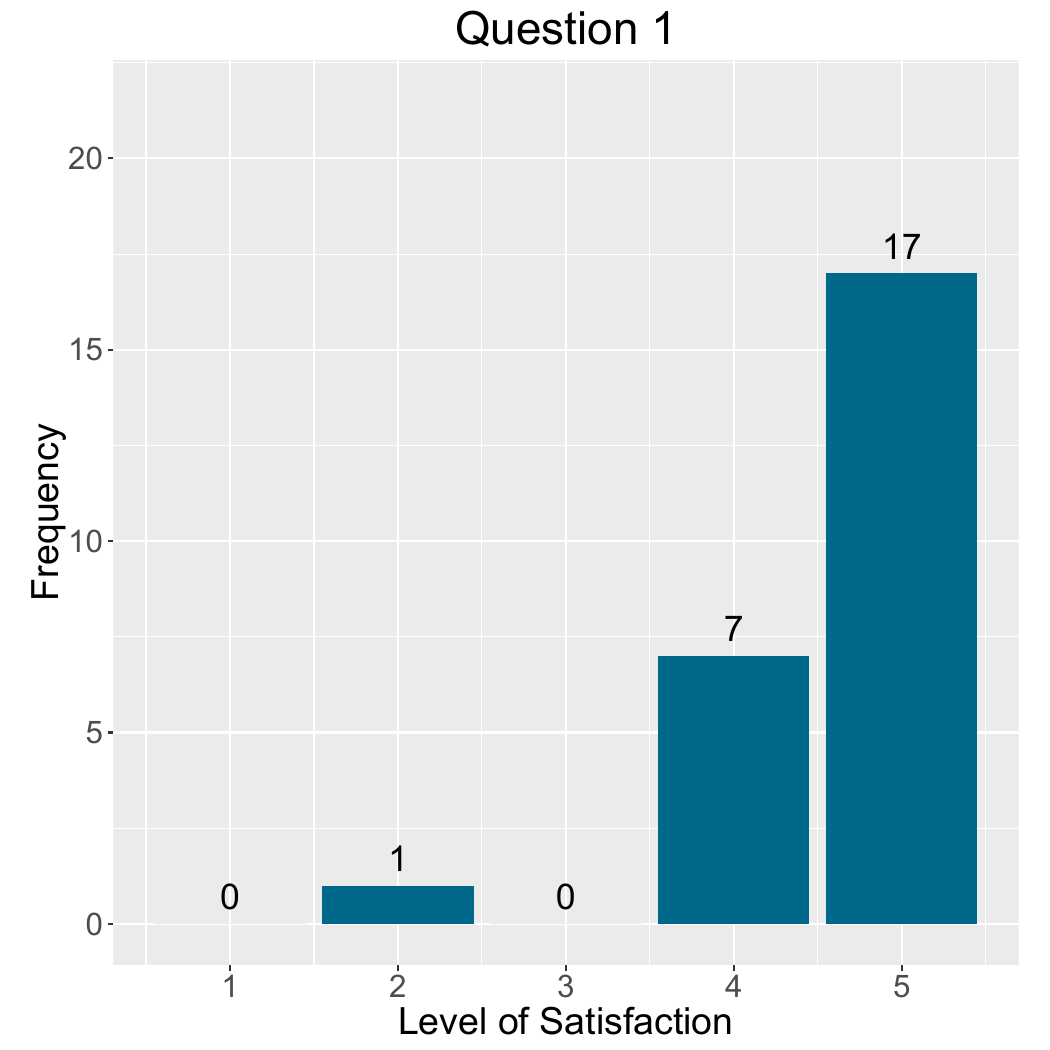} \hspace*{-5mm}
\includegraphics[scale=0.28]{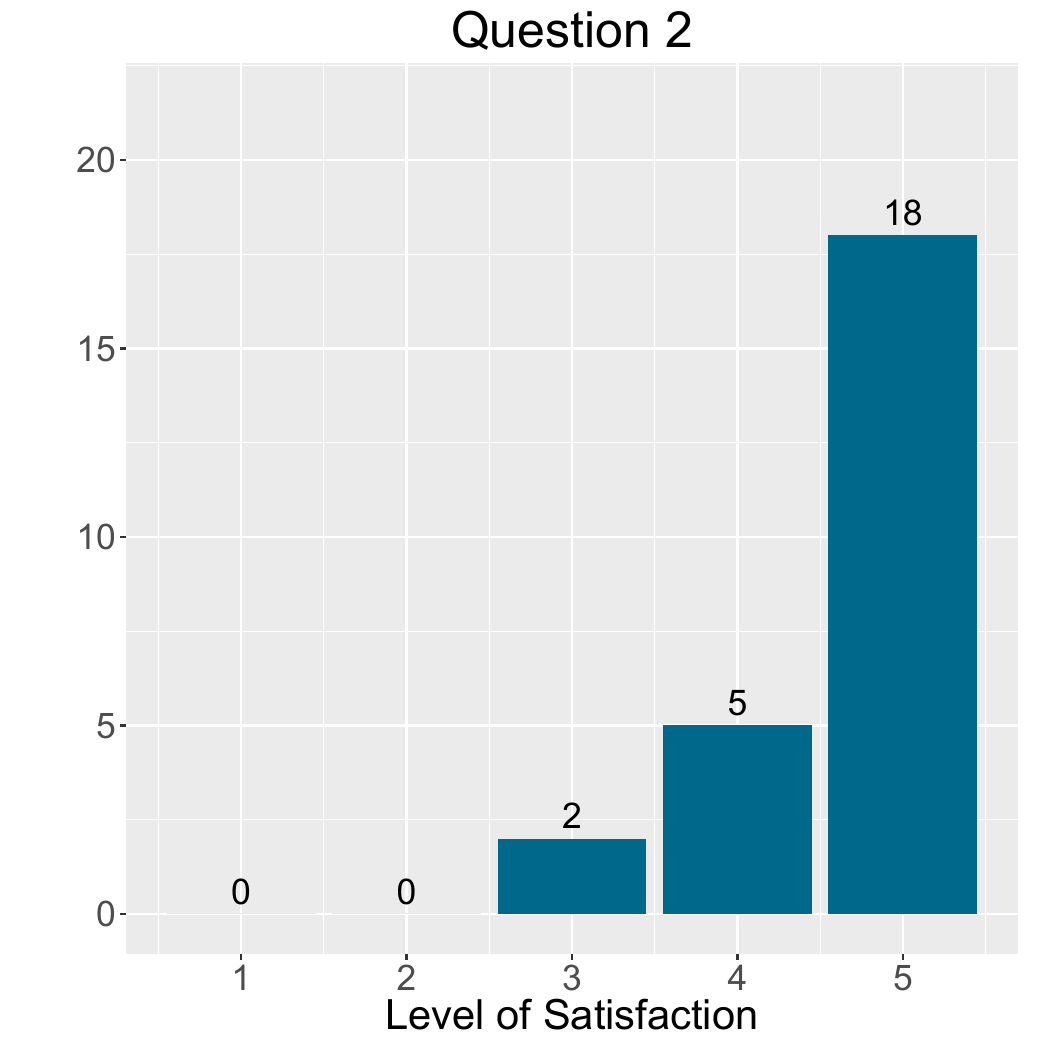}
\hspace*{-5mm}
\includegraphics[scale=0.28]{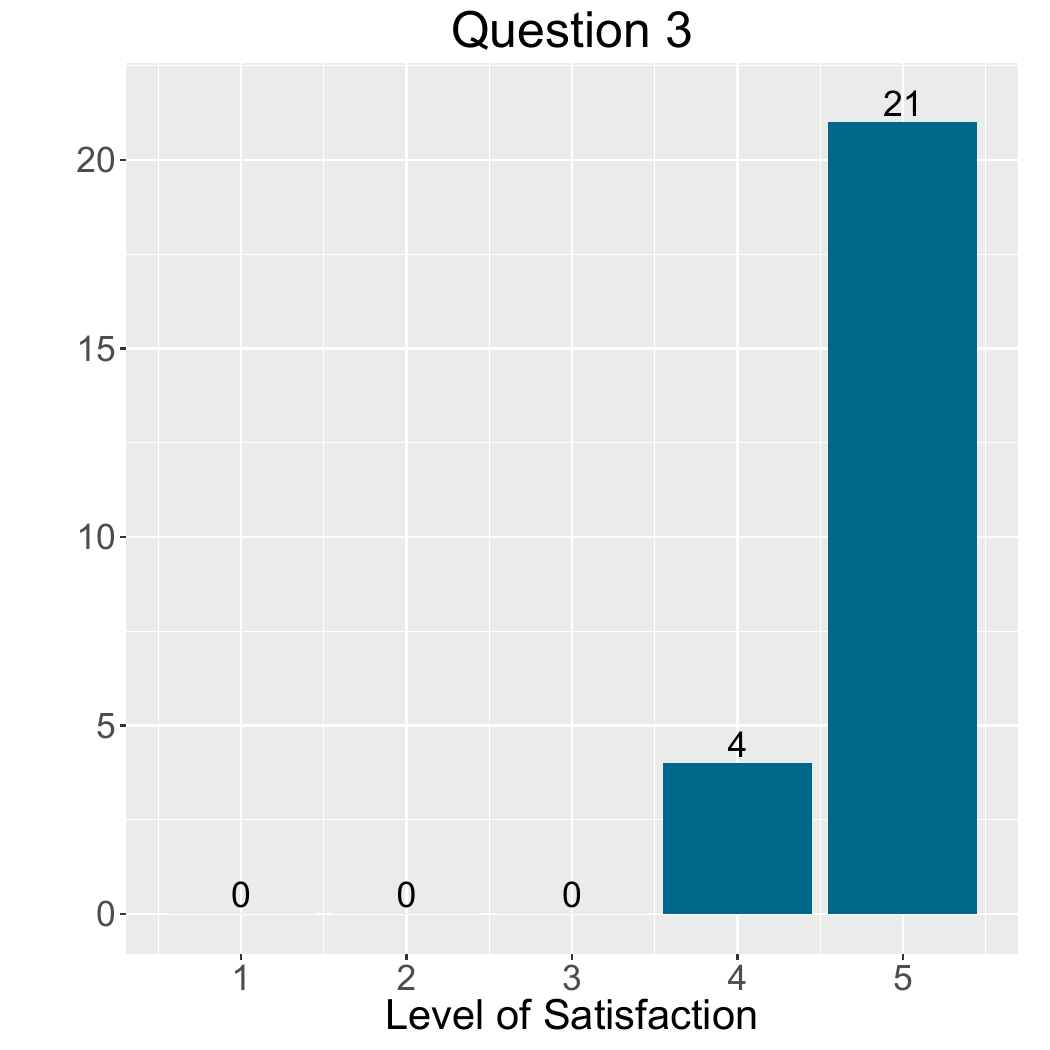}
\caption{Results of a survey conducted with 25 students following a trial of Statverse in a random spatial patterns environment.}
\label{Fig:questions}
\end{figure}

In addition to these questions students were also asked to comment on their familiarity with statistical concepts. The survey was completed by 25 students who fully engaged in the experience. Fig.~\ref{Fig:questions} presents bar plots illustrating the outcomes in each case, summarizing the students' willingness to participate in learning activities that complement and enhance the understanding of complex concepts. 

\section{Statverse in action}

The feedback from the pilot study was crucial, directly influencing the next sprint and refining the Statverse experience for its integration into our programmes. A first course in Probability \& Statistics was launched in August 2024 at \textsc{utfsm}, with \textsc{uoe} also planning a similar initiative for 2025--26.
\subsection{Illustrations}

We have designed and created novel virtual modules focusing on a key concept from each unit of the pilot course: random patterns, probability, normal random variables, central limit theorem, estimation, and hypothesis testing \citep[e.g.,][]{devore2012, blitzstein2019, speegle2021}. For each of these core concepts we have crafted a custom Metaverse teach-in session. Due to space constraints, we focus on briefly highlighting two of these.

\begin{figure}[H]
\centering
\includegraphics[scale=0.4]{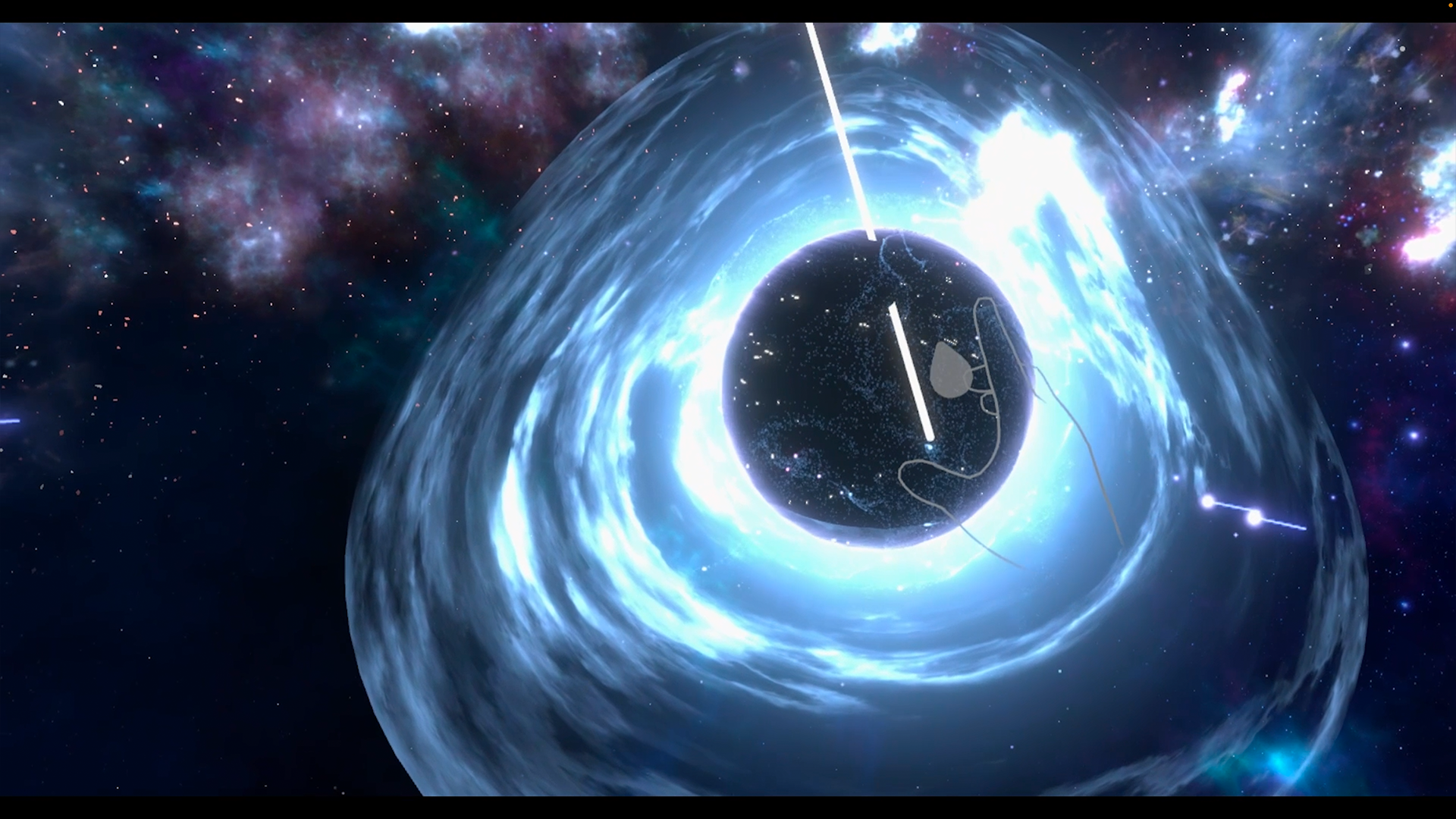}
\includegraphics[scale=0.4]{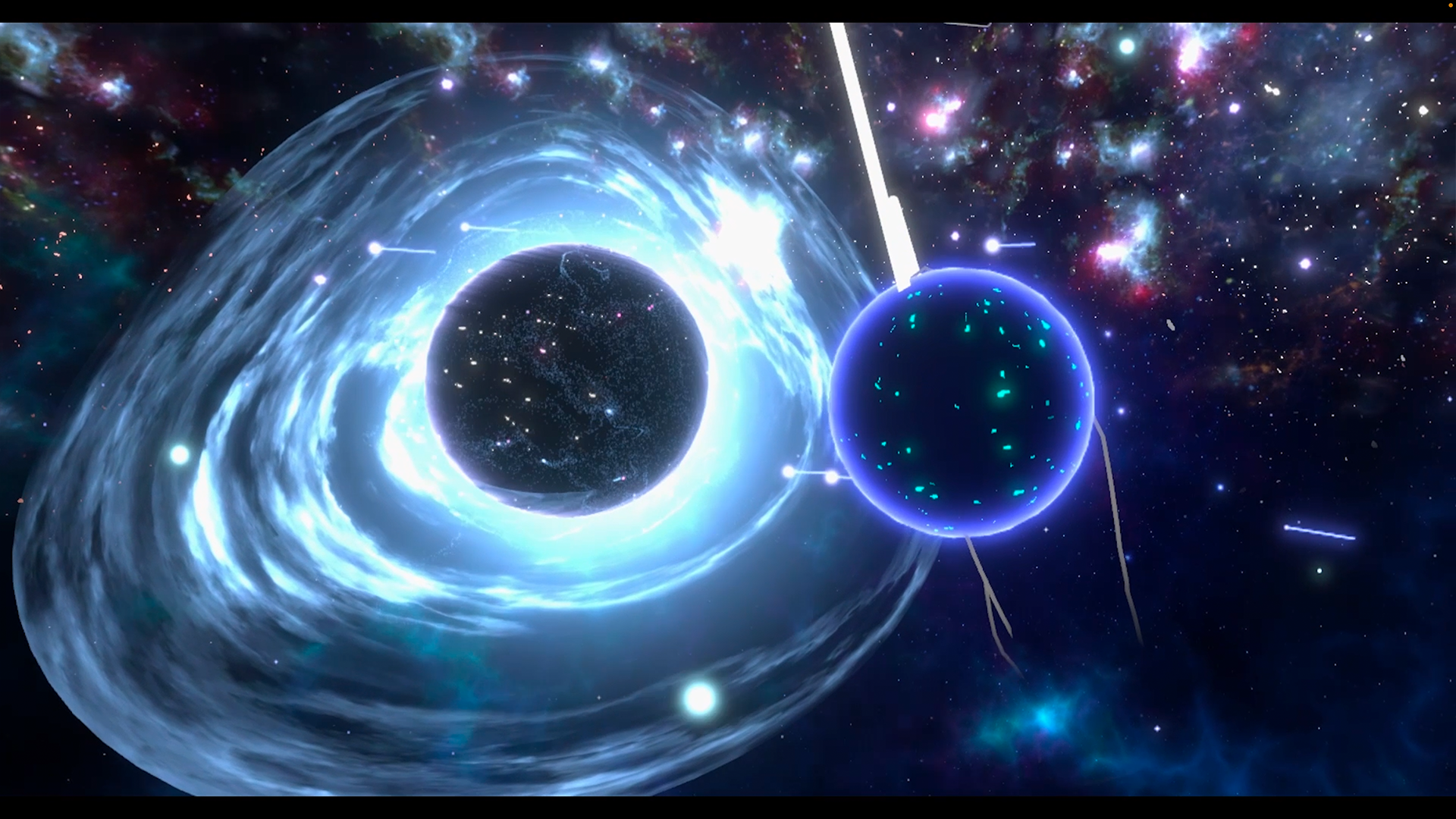}\\
\includegraphics[scale=0.4]{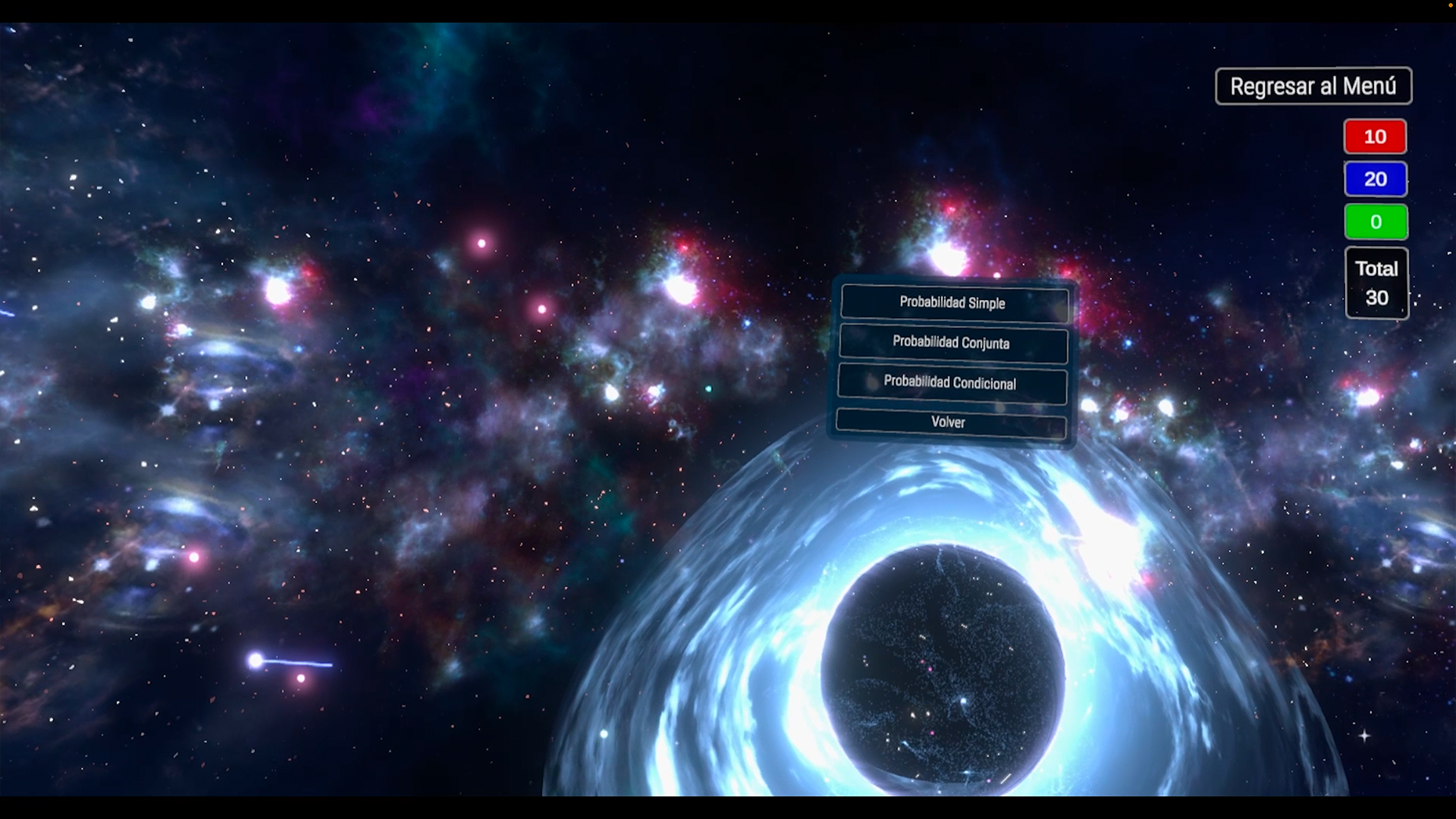}
\includegraphics[scale=0.4]{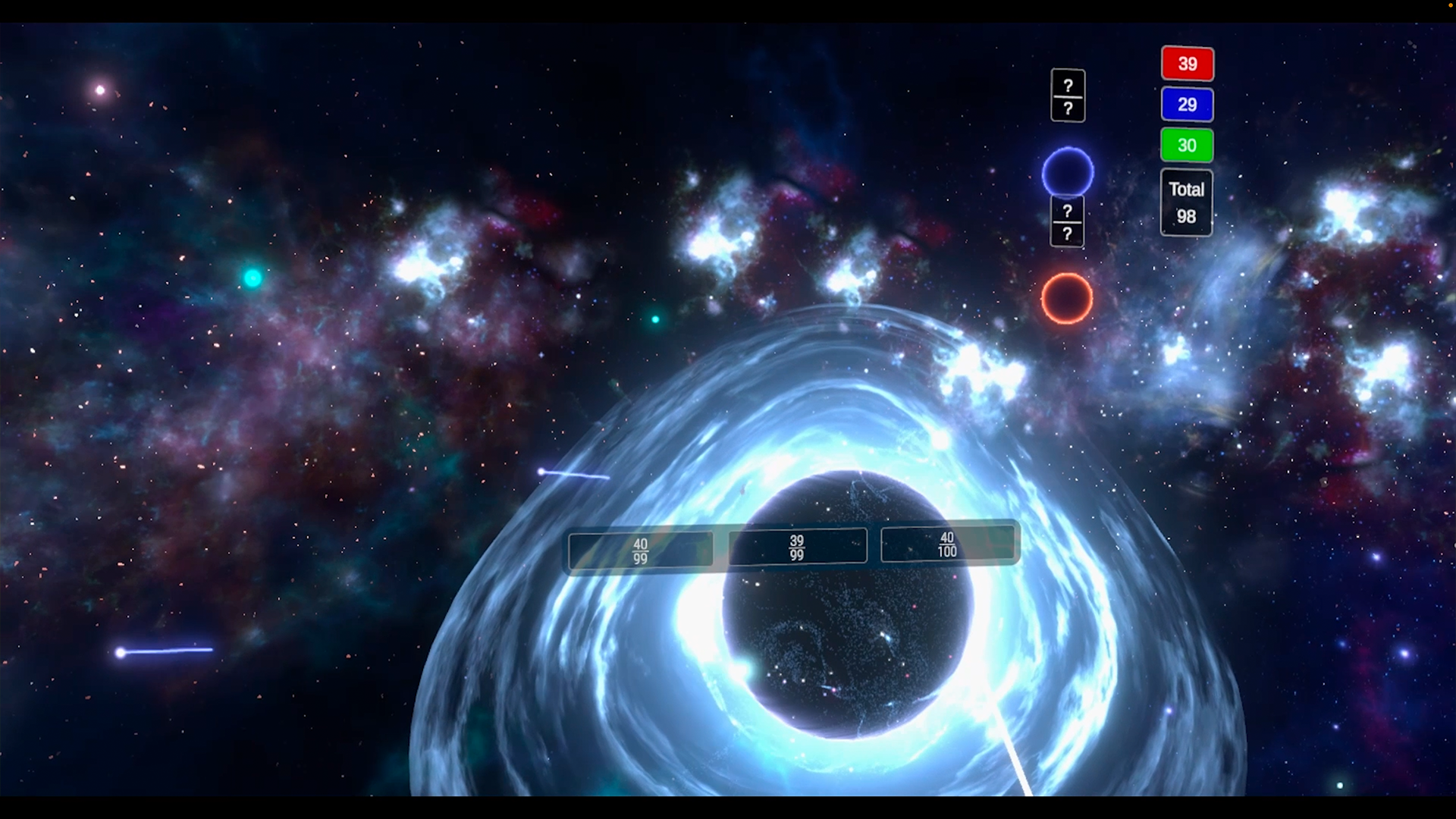}
\caption{Virtual environment of experience 2: Probability teach-in session (\url{https://edin.ac/4ah0Xew}).}
\label{Fig:experience2}
\end{figure}

In Fig.~\ref{Fig:experience2}, we showcase four snapshots from our Virtual Reality module on the notion of probability. Initially, users encounter a simulated black hole (top left) filled with multicolored balls (top right). They can sequentially draw balls to compute probabilities of individual events, intersections, or conditional probabilities (bottom left), by tallying successes against the total ball count (bottom right). The learning section includes exercises with results presented as fractions for simplicity. This setup aims to facilitate learning the concepts of counting measures, sampling without replacement, and probability rules.

\begin{figure}[H]
\centering
\includegraphics[scale=0.32]{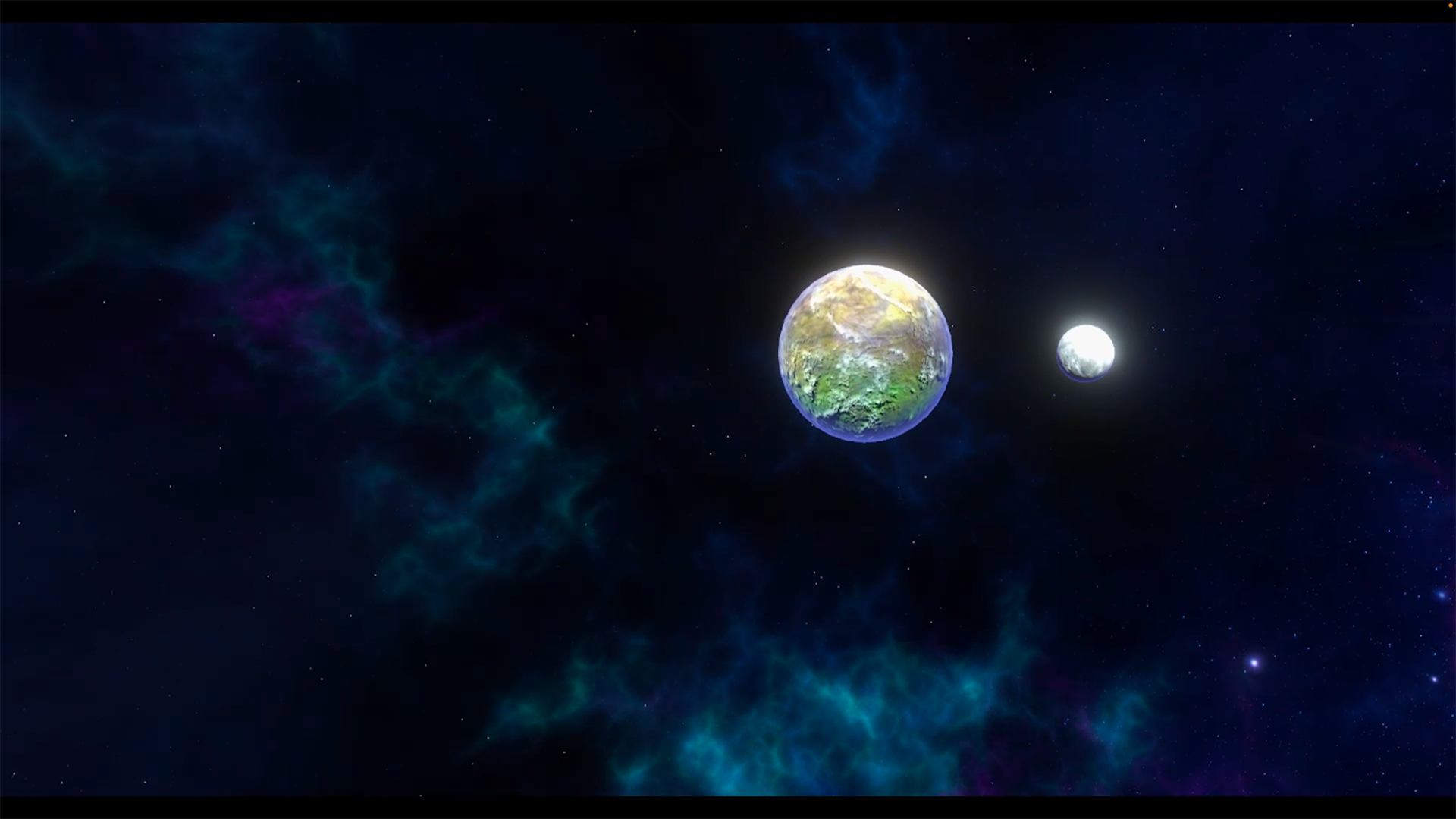}
\includegraphics[scale=0.32]{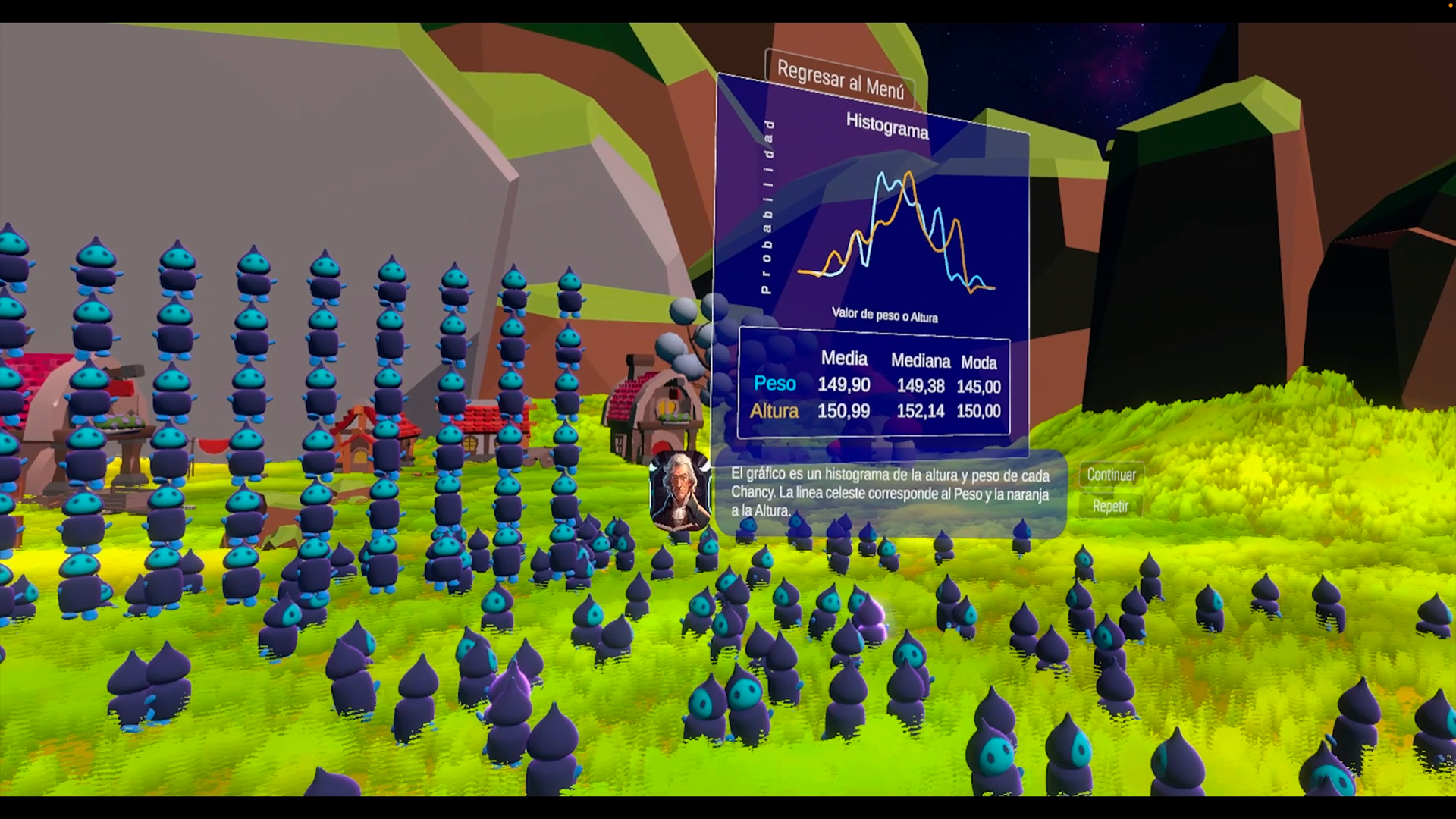}\\
\includegraphics[scale=0.32]{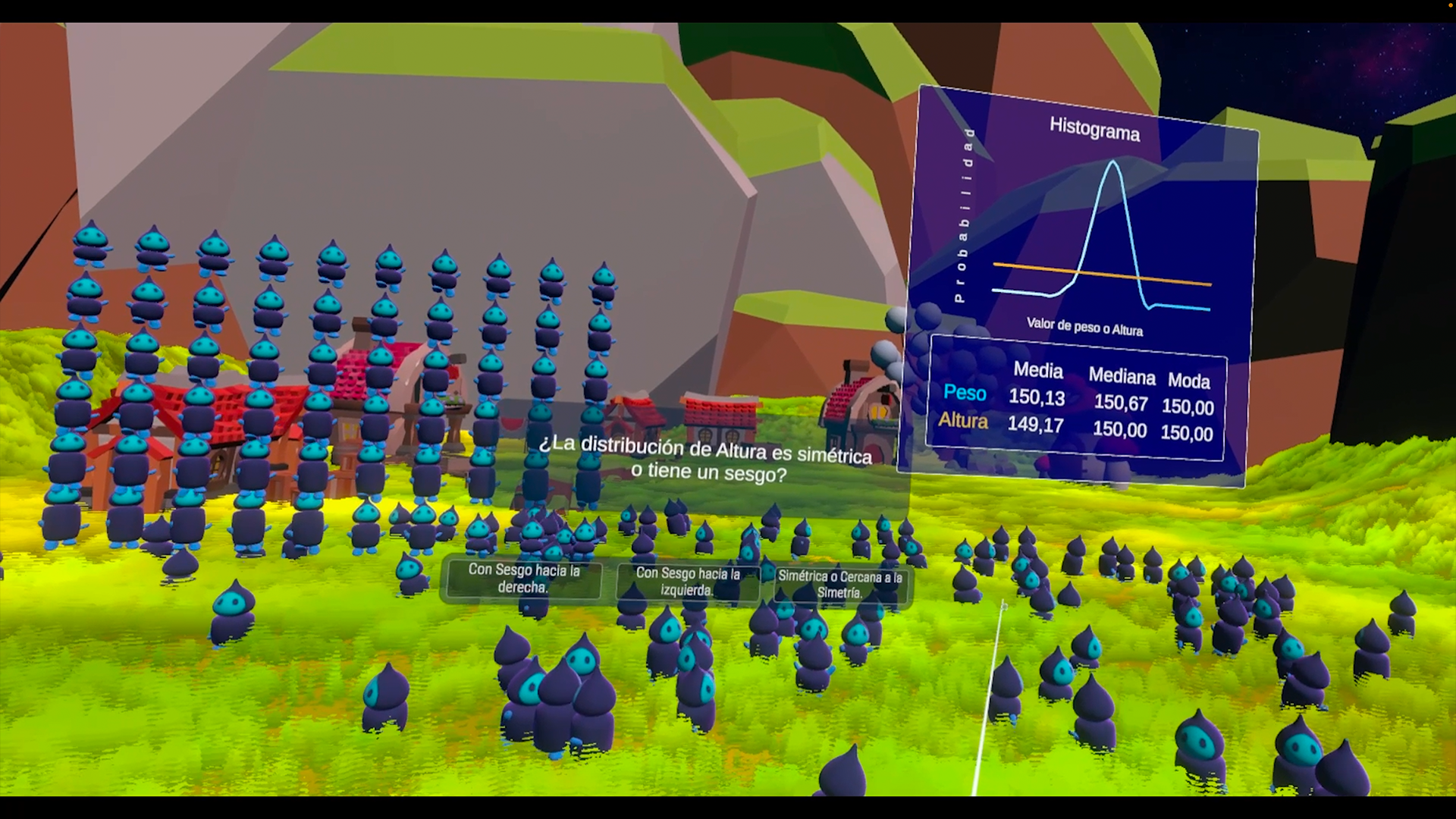}
\includegraphics[scale=0.32]{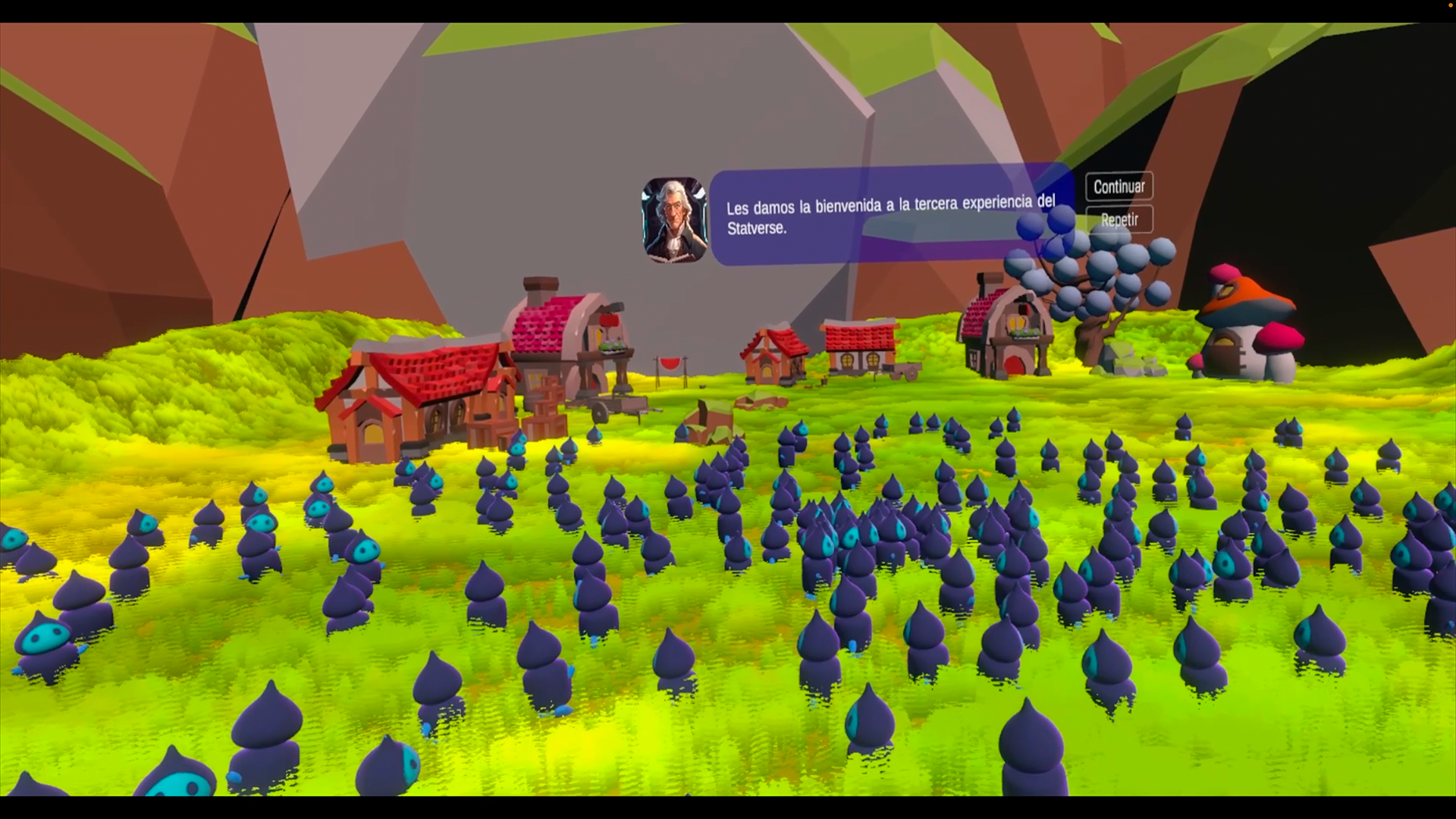}
\caption{Virtual environment of experience 3: The Normal distribution 
 (\url{https://edin.ac/4bleHWM}).\label{Fig:experience3}}
\end{figure}

Next, we introduce the module on data insights. In Fig.~\ref{Fig:experience3}, the user is teleported to a planet (top left) inhabited by Chancys (top right). Chancys vary in height and weight, which are the variables of interest. The user can select as many Chancys as desired, and summary statistics along with a density estimator are displayed on an internal screen. Concepts such as the mean, median, mode, standard deviation, and skewness can be immediately gathered from the information displayed on the screen. While the central limit theorem was not initially planned for this experience, this immersive activity can certainly serve as a starting point for introducing that result by increasing the sample size.

We plan to conduct an expanded study in the Fall semester of 2024, where students will engage in all Statverse activities throughout the term. 

\subsection{Development \& Design}
Some comments on the preparation and implementation of the virtual modules are in order.
We kept in mind the Successive Approximation Model \citep[SAM;][]{allen2012leaving}, aligning it strategically with the methodological framework used by Virtual Reality developers. Following the design phase, the integration of 3D models and interactive elements into the Statverse environment was facilitated using Unity and C\#. Finally, modern principles of Instructional Design \citep[][]{dick2005} were applied to enhance both the framework's effectiveness and the user experience. 

The supplementary materials provide additional insights on the development and implementation of Statverse, which may be helpful for those considering forked versions at their own institution.

\subsection{First Cohort: Feedback \& Assessment}
The first cohort of the new Probability \& Statistics course began in Aug.~2024 with 50 students. Of these, 27 volunteered for Statverse edition, while 23 opted for traditional TA sessions. All students shared the same classroom, but the key difference was that Statverse students engage with the platform, while others attend TA sessions. Both groups used similar materials, However, the TA sessions lacked the visualizations and data tactilizations offered by Statverse within a physical space specifically arranged for the experience, ensuring that movements and displacements are safe and appropriate.

\begin{figure}
\includegraphics[scale = 0.2]{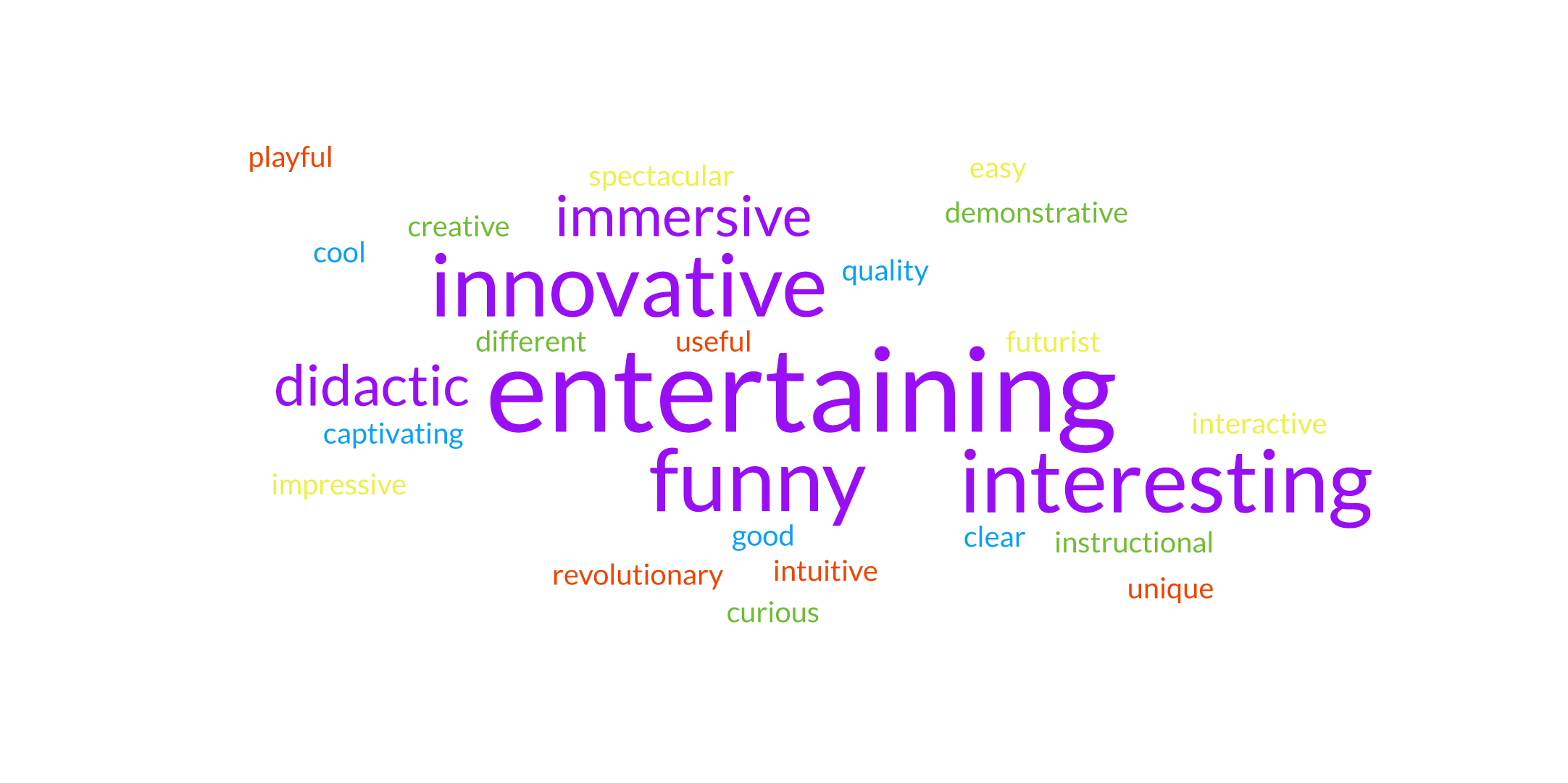} \vspace{-1.5cm}
\caption{\label{wordcloud}Wordcloud of feedback from first cohort.}
\end{figure}

Fig.~\ref{wordcloud} shows a wordcloud based on feedback from the 27 students in the Statverse group, specifically from the virtual environment of the point pattern activity in Fig.~\ref{Fig:patterns}. The most frequent words in the feedback include `entertaining,' `interesting,' and `innovative,' as seen in Fig.~\ref{wordcloud}.

Next, we move to assessment.  All students participated in the first midterm and one of questions in the exam involved aspects related with representation and visualization of point patterns:
\begin{quote}\it 
``Plot a regular pattern in a square with finite sides on the plane, containing 10 points."
\end{quote}
Random, aggregated, and regular patterns are well-known concepts in the point process literature \citep{schabenberger2017}. Performance was analyzed keeping in mind students' prior engagement with experience 1 (Fig.~\ref{Fig:patterns}). The results were as follows:
\begin{itemize}
\item Of the 27 students who completed experience 1, 25 achieved full marks on the question, corresponding to 92.6\%.
\item Among the 23 students who did not engage in experience 1, 11 achieved full marks on the question, corresponding to 47.8\%.
\item Out of the total 50 students who took the midterm, 36 achieved full marks on the question, representing 72\%. (Overall performance).
\end{itemize}
Hence, although both groups had similar preparation in theory, the group that interacted with Statverse performed better on this question.

\subsection{Learning Outcomes, Competencies \& Enduring Understandings}

Two learning outcomes of the Probability and Statistics course, which align with the impact that Statverse can deliver, are as follows: organizing and summarizing data effectively in engineering applications while interpreting it appropriately, and developing implementations of statistical procedures with specialized software support. Within this framework,  instructional design was structured across three stages: learning, practice, and evaluation, ensuring the long-term retention of fundamental statistical concepts. For example, in experiences 1 to 3, the instructional design aimed to facilitate students' recognition and identification of point patterns, enhance their understanding of marginal, joint, and conditional probabilities, and improve their ability to analyze probability distributions and statistical estimates.

Beyond the scope of this course, research has demonstrated that immersion in a digital environment can enhance education through at least three mechanisms: enabling multiple perspectives, facilitating situated learning, and enhancing transferability \citep{Dede2009}.

A presentation delivered at the QS Reimagine Education 2025 conference\footnote{https://www.reimagine-education.com}, held in London in December 2025, indicates that Statverse has been well received by the community and aligns with the innovation trends promoted by the organizers. This suggests a strong potential impact among educators seeking innovative approaches for courses that have traditionally challenged students, with possible benefits for student retention.

\section{Closing remarks}
Data Science stands to benefit significantly from the rapid advancements in the Metaverse and Statistical Computing. Although these fields are still in their infancy, it is crucial for Data Science to begin exploring how it can integrate with these emerging paradigms at this early stage. 

Statverse provides a initial approach to integrate these emerging technologies. Our main goal here was to document our progress on interfacing Data Science and the Metaverse and invite others to integrate similar strategies into their programs.  While the preliminary findings are encouraging, we openly acknowledge that this is a pioneer project and a more thorough analysis will be conducted once we gather more data from future cohorts.

The \textsc{utfsm}--\textsc{uoe} implementation is merely one instance of this concept in action. There is still a lot of work to be done, and we encourage others to create their own adaptations of this concept---embracing and integrating Virtual Reality and Spatial Computing into the Data Science classroom. The quest to interface Data Science with Virtual Reality is not new \citep{cook2001}, yet the industry is now advancing at an unprecedented pace, suggesting that many of these developments may soon materialize.  Clearly, key references in the field \citep[e.g.,][]{gelman2017} provide limited guidance on the challenges and opportunities posed by these technological advances.

Data visualization and data tactilization are at the heart of Statverse's approach, transforming abstract data concepts into interactive and immersive experiences. While the focus above has been placed on education, Data Science research will greatly benefit from embracing the same paradigm, as well as from putting even further emphasis on Metaverse and Spatial Computing. While some exceptions exist \citep[e.g.,][]{castruccio2019}, more work could be developed on that front to enhance interactive and immersive tactile environments as well as to foster engaging visualizations.

\subsection*{Disclosure Statement}
The authors have no conflicts of interest to declare.

\subsection*{Confidentiality of Respondents}
Participant identities were not collected and confidentiality was maintained throughout the survey reported in Section~\ref{unveiled}.

\subsection*{Acknowledgments}
Funding from the Royal Society of Edinburgh is gratefully acknowledged. The authors are grateful to Andrés Fuentes,  Felipe Osorio, Rodrigo Díaz, Hernán Rojas,   and Clemente Ferrer for helpful discussions and support. This work  was funded by Universidad Técnica Federico Santa María through VRA, DTD, DEA, and DIRED. We also acknowledge the support from Project 2030, led by Mario Toledo. 

\nocite{*}
\bibliography{references}

\begin{thebibliography}{99}
\bibitem[Bates(2019)]{Bates:2019} 
Bates, T. (2019). Teaching in a Digital Age: Guidelines for Designing Teaching and Learning, 2nd Edition. Vancouver: Tony Bates Ltd.

\bibitem[Morrison(2010)]{Morrison:2010} 
Morrison, G. R. (2010). Designing Effective Instruction, 6th Edition. New York: John Wiley \& Sons. 
\end{thebibliography}
\bibliographystyle{plainnat}

\newpage
\begin{center}
\section*{\large Supplementary Materials for: \\
"Welcome to the Statverse: A Metaverse of Data Science"}
\end{center}

In this web appendix, we provide additional practical information that may be helpful for creating tools similar to Statverse. In developing and implementing Statverse, we followed various approaches from the ADDIE model \citep{Morrison:2010, Bates:2019}, commonly used for designing e-learning courses. We believe these steps could assist others considering forked versions of Statverse at their own institutions. The steps are as follows:

\begin{itemize}
\item[1.] {\bf Draft the Project}. The first stage involves drafting a project with clear objectives that justifies the need for an educational product using virtual reality technology. In our case we had to make sure the project aligned with the institution's strategic plans to secure budgetary funds for this educational technology.

\item[2.] {\bf Conceptual Design}. When designing a new physical or digital product, it is advisable to start with a conceptual design. This subprocess guides the creation from identified needs, through user studies and brainstorming, to the first draft of the final product. 
  The conceptual design should consider the user experience, proposing ideas on how students will interact with the product to efficiently achieve learning objectives.

\item[3.] {\bf Technical Team}. There are two possibilities: either establish an internal team for virtual reality development or hire a specialized company in this field. We recommend the second option, especially for those starting out, as an external company brings a fresh perspective to the conceptual design, generating new ideas that may not have been considered. The first approach should be avoided if there is no prior experience in such projects, as assembling the necessary team (Scrum Master, Unity or Unreal developer, QA, methodological advisor, testers, 3D designer, and texture artist) and acquiring the appropriate technological equipment (gaming PC, VR headsets, link cables, etc.) are complex and costly tasks.

\item[4.] {\bf Product Development}. The hired company was responsible for overseeing the digital product development process until its completion. It is recommended to monitor progress through weekly meetings for iterative adjustments and to ensure the final product meets client requirements. 
  In our case the process began with detailed engineering, defining aspects such as colors, environment, avatar, texts, interactions, and functionalities, followed by environment development and programming. Finally, after multiple reviews and agreements, the first functional prototype was achieved.

\item[5.] {\bf Implementation}. With the product properly tested and approved, it should be implemented in the course, ensuring the following critical aspects: schedule, space, and technology. The schedule for students to engage in the virtual reality activity must be coordinated within the course schedule. A physical space or laboratory should be allocated within the institution for students to experience virtual reality safely and seamlessly, ideally a space of at least 2x2 meters without obstacles. The technology should be operational with supervision to ensure the proper functioning of the final product.

\item[6.] {\bf Results \& Achievements}. The ultimate goal should always be to achieve the learning objectives for which the product was designed. This can be measured through the product's outcomes or student performance, using written and oral assessments where students answer conceptual questions and contemplate what they have learned. These results should be designed using appropriate statistical  models, such  that the main findings can be used for scientific research, press releases, and report development.
\end{itemize}

\end{document}